\documentclass[preprint,aps]{revtex4}
\usepackage{graphicx}
\usepackage{booktabs}
\usepackage{xcolor}
\usepackage{hyperref}
\usepackage{multirow}
\usepackage{setspace}

\newcommand{\comment}[1]{}

\begin{document}

\title{Predicting economic market crises using measures of collective panic}
\date{August 26, 2010; revised February 13, 2011}

\author{Dion Harmon$^1$, Marcus A. M. de Aguiar$^{1,2}$,
David D. Chinellato$^{1,2}$, Dan Braha$^{1,3}$, Irving R.
Epstein$^{1,4}$, Yaneer Bar-Yam$^1$}

\affiliation{$^1$New England Complex Systems Institute,  238 Main St.,
  Suite 319, Cambridge, MA 02142, USA \\ $^2$Universidade Estadual
  de Campinas, Unicamp 13083-970,
  Campinas, SP, Brasil\\ $^3$ University of Massachusetts Dartmouth,
  Dartmouth, MA 02747\\ $^4$ Brandeis University, MS 015 Waltham MA, 02454}

\begin{abstract}
Predicting panic is of critical importance in many areas of human and
animal behavior, notably in the context of economics. The recent
financial crisis is a case in point. Panic may be due to a specific
external threat, or self-generated nervousness. Here we show that the
recent economic crisis and earlier large single-day panics were preceded by
extended periods of high levels of market mimicry --- direct evidence
of uncertainty and nervousness, and of the comparatively weak influence
of external news. High levels of mimicry can be a quite general
indicator of the potential for self-organized crises.
\end{abstract}

\maketitle

In sociology \cite{wolfenstein,smelser,quarantelli,mawson}, panic has
been defined as a collective flight from a real or imagined threat. In
economics, bank runs occur at least in part because of the risk to the
individual from the bank run itself---and may be triggered by
predisposing conditions, external (perhaps catastrophic) events, or
even randomly \cite{diamonddybvig,calomirisgorton1990}. Although
empirical studies of panic are difficult
\cite{mannnageldowling,galbraith,kindleberger}, efforts to
distinguish endogenous (self-generated) and exogenous market panics
from oscillations of market indices have met with some success
\cite{sornette8,feigenbaum1996,sornette7,stauf,sornette-book}, though
the conclusions have been debated
\cite{feigenbaum2,sornette4,bree,cho}.
Market behavior is often considered to reflect external economic news,
though empirical evidence has been presented to challenge this
connection \cite{cutlerpoterbasummers}. Efforts to characterize events
range from the Hindenburg Omen \cite{morris} to microdynamic models
\cite{solomon} and to the demonstration that market behaviors are
invariant across many scales \cite{stanley}. Other work has looked at
relationships of market behavior with internet search
\cite{preisstanley}. Panic can be considered a `critical transition' for which
early warnings are being sought \cite{scheffer}. 
The ``collective flight" aspect of such a transition should be revealed in
measures of mimicry that is considered central to panic. Here
we use co-movement data to evaluate whether the recent market crisis
and earlier one-day crashes are internally generated or externally
triggered. Based upon a hypothesis about mimicry, we construct a model
that includes both mimicry and external factors and test it empirically
against the daily extent of co-movement. Our objective is to determine
the relative importance of internal and external causes, and, where
internal causes are important, to find a signature of self-induced
panic, which can be used to predict panic.

The literature generally uses volatility and the correlation between 
stock prices to characterize risk.\cite{ross,chamberlain,mantegna,onnela,smith,text,marketnets}
These measures are sensitive to the magnitude of price movement and therefore increase 
dramatically when there is a market crash. Studies find that, on average, volatility increases 
following price declines, but do not show higher volatility is followed by price declines.\cite{black,christie,nelson,bekaert,wu} 
We are interested in the extent to which stocks move together. The extent of such co-movement 
may be large even when price movements are small. Indeed, even when 
price changes are small, we expect that co-movement itself is the collective 
behavior that is characteristic of panic, or panicky behavior that precedes a panic. Thus, rather
than measuring volatility or correlation, we measure the fraction 
of stocks that move in the same direction. We find that this increases 
well before the market crash, and there is significant advance warning to 
provide a clear indicator of an impending crash. The existence of the indicator shows that 
market crashes are preceded by nervousness that gives rise to following 
behavior --- increased collective behavior prior to a panic. 

We consider the ``co-movement" of stocks over time by plotting the
number of days in a year that a particular fraction of the market moves up
(the complement moving down). Intuitively, if substantially more or
less than 50\% of the market moves in the same direction, this
represents co-movement. As shown in Fig. 1, the results indicate that
in 2000, the curve is peaked near 1/2, so that approximately 50\% of
stocks are moving up or down on any given day. Over the decade of the
2000s, however, the curve became progressively flatter---in 2008 the
likelihood of any fraction is almost the same for any value. The
probability that a large fraction of the market moves in the same
direction, either up or down, on any given day, increased dramatically.
Such high levels of co-movement may manifest the collective behavior we
are searching for.

\begin{figure}
  \sf
  \centering
  \includegraphics[height=6in]{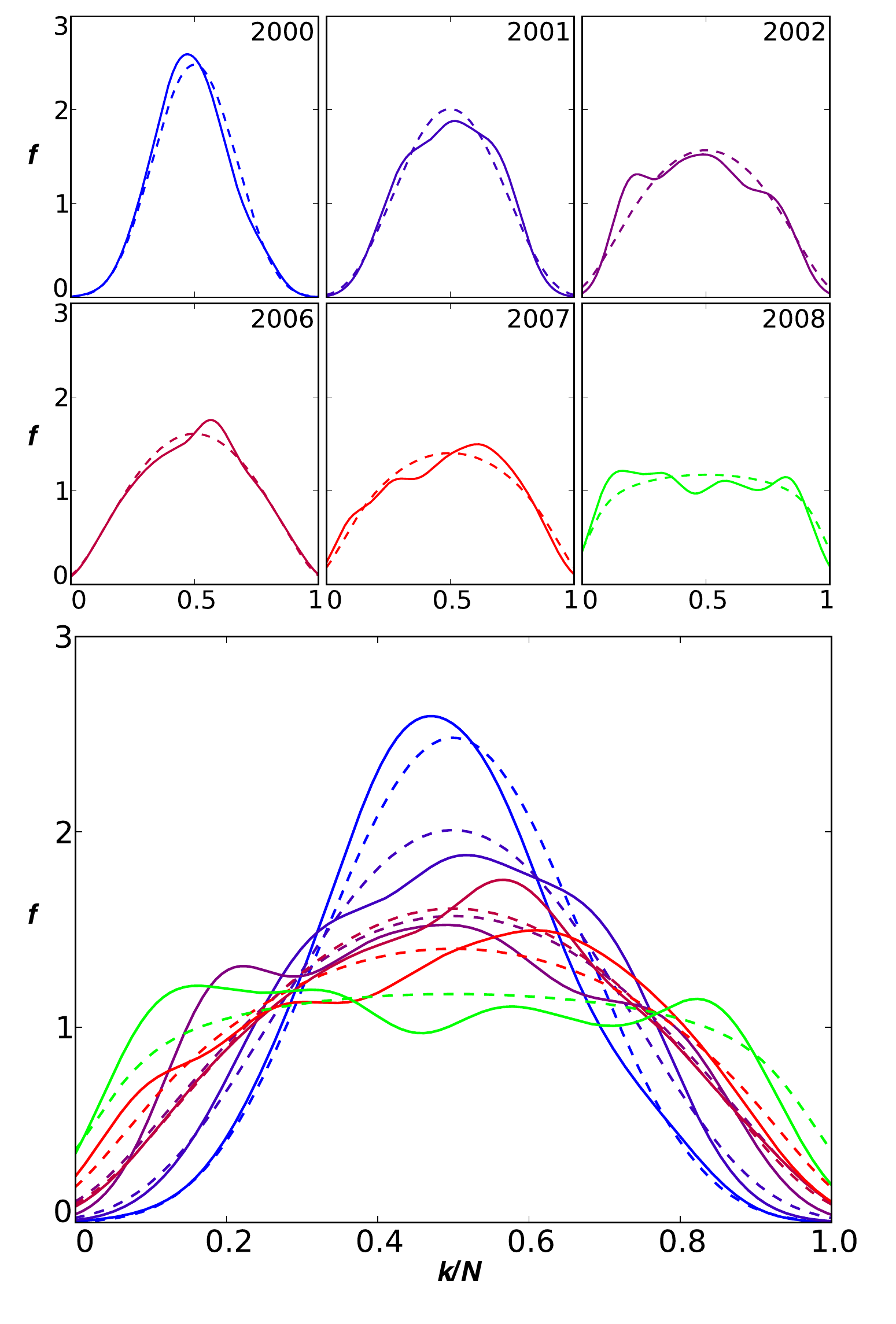}
  \caption{The co-movement of stocks. Plotted is the fraction of trading days during the year ($f$, vertical axis)
   in which a certain fraction of stocks ($k/N$, horizontal axis) moved up.
   Empirical data are shown (solid lines) along with one-parameter theoretical fits
   (dashed lines) for the years indicated. Three years are omitted that do not differ
   much from the year immediately preceding and after them.  Bottom panel combines
   all of the years shown.
   Stocks included are from the Russell~3000 that trade on the NYSE or Nasdaq.
   Curves are kernel density estimates with Gaussian kernels ($\sigma=0.06$).  Fits
    pass the $\chi^2$ goodness-of-fit test (the deviation of the data from the theoretical
    distribution is not statistically significant at the 25\% level).}
\label{fig:melt_panels}
\end{figure}

To quantitatively describe co-movement, we start from a 
behavioral economics model of a single stock that describes trend-following
``bandwagons." It has been shown that investors can benefit from
trend-following \cite{delong,delong2,scharf,welch}. Moreover, there is no
need for the change to be based upon fundamental value for it to
provide benefit to the investors \cite{delong,delong2}. When individuals observe
that a stock increases (decreases) in value, and choose to buy (sell)
in anticipation of future increases (decreases), this self-consistently
generates the desired direction of change. Such a ``bandwagon" effect can
undermine the assumptions of market equilibrium. We hypothesize that
this trend-following mimicry across multiple stocks can cause a
marketwide panic, and we build a model to capture its signature. We
assume that investors in a stock observe three things, the direction of
their stock, external indicators of the economy, and the direction of
other stocks. The last of these is the potential origin of self-induced
market-wide panic.

To model the co-movement fraction, we represent only whether a stock value rises or falls.
This enables us to directly characterize the degree to which stocks move together and not how far
they move at any particular time.
Stocks are represented by nodes of a network and influences between
stocks by links between nodes, an appropriate representation for market
analysis \cite{schweitzer,mantegna,onnela,smith}. We consider both
fully and partly connected networks. Every day, each of the $N$ nodes
is labeled by a sign $({+}/{-})$ indicating the daily return of the
stock. Market dynamics is simulated by randomly selecting nodes, which
maintain their current sign or randomly copy the sign of one of their
connected neighbors. To represent external influences, we add nodes
that influence others, but are not themselves influenced, i.e.
``fixed'' nodes. The number of fixed nodes influencing in a positive
direction is $U$ and the number influencing in a negative direction is
$D$.  The effective strength of the positive and negative external
influences is given by the number of these nodes. External influences
of opposite types do not cancel; instead larger $U$ and $D$ reflect
increasing probability that external influences determine the returns
of a stock independent of the changes in other stocks. This is the
conventional view that news is responsible for the market behavior.
Good news would be represented by $U$ greater than $D$, bad news by $D$
greater than $U$.

We have previously proposed this model as a widely applicable theory of collective behavior
of complex systems.\cite{chinellato} Successful matching to data will be a confirmation
of this theory. It has also been previously identified as a model of conformity
and non-conformity in social systems,\cite{arthur} and it has been studied 
in application to evolutionary dynamics.\cite{ewens} 

This model can be solved exactly for a fully connected network (see
Appendix). We obtain the probability of a co-movement fraction:
\begin{equation}
  f(k/N)=\frac{
    {U+k-1\choose k}
    {N+D-k-1 \choose N-k}}{
    {N+D+U-1 \choose N}}
\end{equation}
where N is the number of stocks, $k$ is the number of stocks with positive returns and ${n\choose
k}$ are binomial coefficients. The behavior is controlled by the strength
of external stimuli, U and D, compared to the strength of interactions
within the network, and the relative bias of the external influence
toward positive, U, or negative, D, effects. When interactions are weak compared to
external forces ($D, ~U >> 1$), the distribution is essentially
normal. When internal interactions are strong (small $D$, $U$), the
distribution is neither normal nor long-tailed. Instead it becomes flatter,
becoming exactly flat at the critical value ($D=U=1$), where the external
influences only have the strength of a single node. Analytic
continuation allows $U$ and $D$ to be extended to non-integer values.
There are three parameters of the distribution, $D$, $U$, $N$, but the third is fixed
to the number of stocks. We can compare this to the binomial or normal
distributions, which are specified by two parameters, the average and
standard deviation. The distribution we obtain has a wider
range of behaviors, and the normal distribution arises as a limiting case.

If we consider a more complete model of influences, in which investors
of one stock only consider specific other stocks as guides, we have a
partly connected network. We have studied the dynamics of such networks
analytically and through simulations, and the primary modification from
fully-connected networks is to amplify the effect of the external
influences (see Appendix). As the links within the
network are fewer, the network can be approximated by a more weakly
coupled, fully connected network, with a weakening factor given by the
average number of links compared to the number of possible links.
Similarly, if only a subset of the external influences are considered
relevant for the return of a specific stock, the relative strength of
the external influences can be replaced by weaker, uniform external
influences. Otherwise, for many cases, the shape of the distribution is
not significantly affected. The model thus measures the relative
strengths of the internal and external influences rather than the
absolute strength of either. The model's robustness indicates a
universality across a wide range of network topologies, suggesting
applicability to real world systems.

Compared with recent empirical market data in Fig. 1,
the model fits remarkably well. A Gaussian model fits the early 
years, less well in the final years, and does not fit 2008. 
The good agreement of our model is obtained
with equal up and down influences, $U=D$, which is the only
adjustable parameter.
This implies that whether the market value is trending up or down, or has
large one day drops, over a period of a year co-movements occur symmetrically
in both up and down directions.
Parameter values are given in Table~\ref{table:fit-values}.

\begin{table}[ht] \begin{center} \sf
    \begin{tabular}{cc}
  \toprule[0.3mm]
\multicolumn{1}{c}{Year} &
  \multicolumn{1}{c}{$ U = D$}\\
\midrule[0.2mm]
2000  &  $5.79$ \\
2001  &  $3.66$ \\
2002  & $2.21$ \\
2006  & $2.32$ \\
2007  & $1.77$ \\
2008  & $1.24$ \\
\bottomrule[0.3mm]
 \end{tabular}
 \caption{Model parameter values used to generate the distributions in
   Figure~\ref{fig:melt_panels}. Empirically, we find that stock
return distributions are symmetric, reducing our model to only one free
parameter, $D=U$.  Similar results are obtained using direct fits and by
using the standard deviation of the distribution (see Appendix).}
 \label{table:fit-values}
\end{center}
\end{table}
The economic crisis period's flat distribution corresponds to $D=U=1$.
This is the critical value of the model where external influences are very weak compared
to the influences among stocks as a whole. By contrast, predominantly negative
effects, $D>U$, would manifest as a distribution whose mean is shifted to the left.
Thus, rather than negative news, uncertainty and collective mimicry led to a
self-organized crash.

\begin{figure}
\sf
\centering
\includegraphics[height=3in]{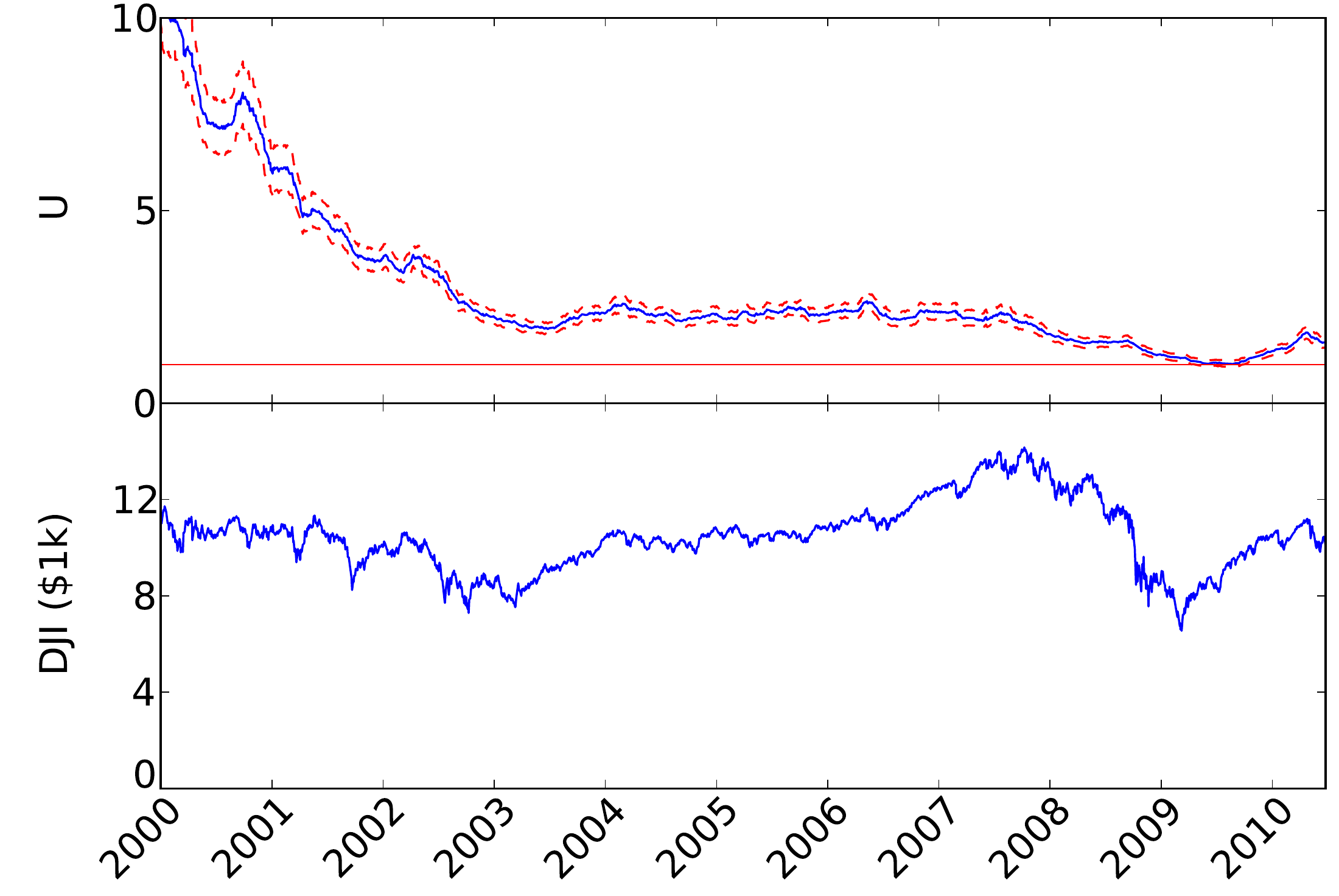}
\caption{Model parameter (top panel) and the Dow Jones Industrial
  Average (bottom panel) for the period 2000-2010. Estimates of the model parameter are shown
  at the end of the year-long period for which $\mathsf U$ was
  estimated.  Sampling error estimates are drawn at~$\mathsf \pm 1$
  standard deviations. Positive-return distributions are computed from
  the daily returns of stocks of the Russell~3000.}
  \label{fig:parameter-short}
\end{figure}

The flattening of the stock market distribution may serve as a measure
of market vulnerability to panic, and the
projection of a flat distribution observed in the economic crisis can
be used as an early warning signal.
Figure~\ref{fig:parameter-short} shows the empirical results of the single parameter from
2000--2010.  We note that the average used for the value at any point
of time is from the period of~12 months prior to that time in order to
evaluate the predictive ability.  A significant drop occurred in the
2000--2002 period, followed by a plateau that declined gradually
beginning in mid-2007 until it hit the critical value at $U=1$.  This
suggests the market was vulnerable well before the financial crisis,
and the gradual decrease before the crisis suggests that the crisis
could have been anticipated.

\begin{figure}
\sf
\centering
\includegraphics[height=4in]{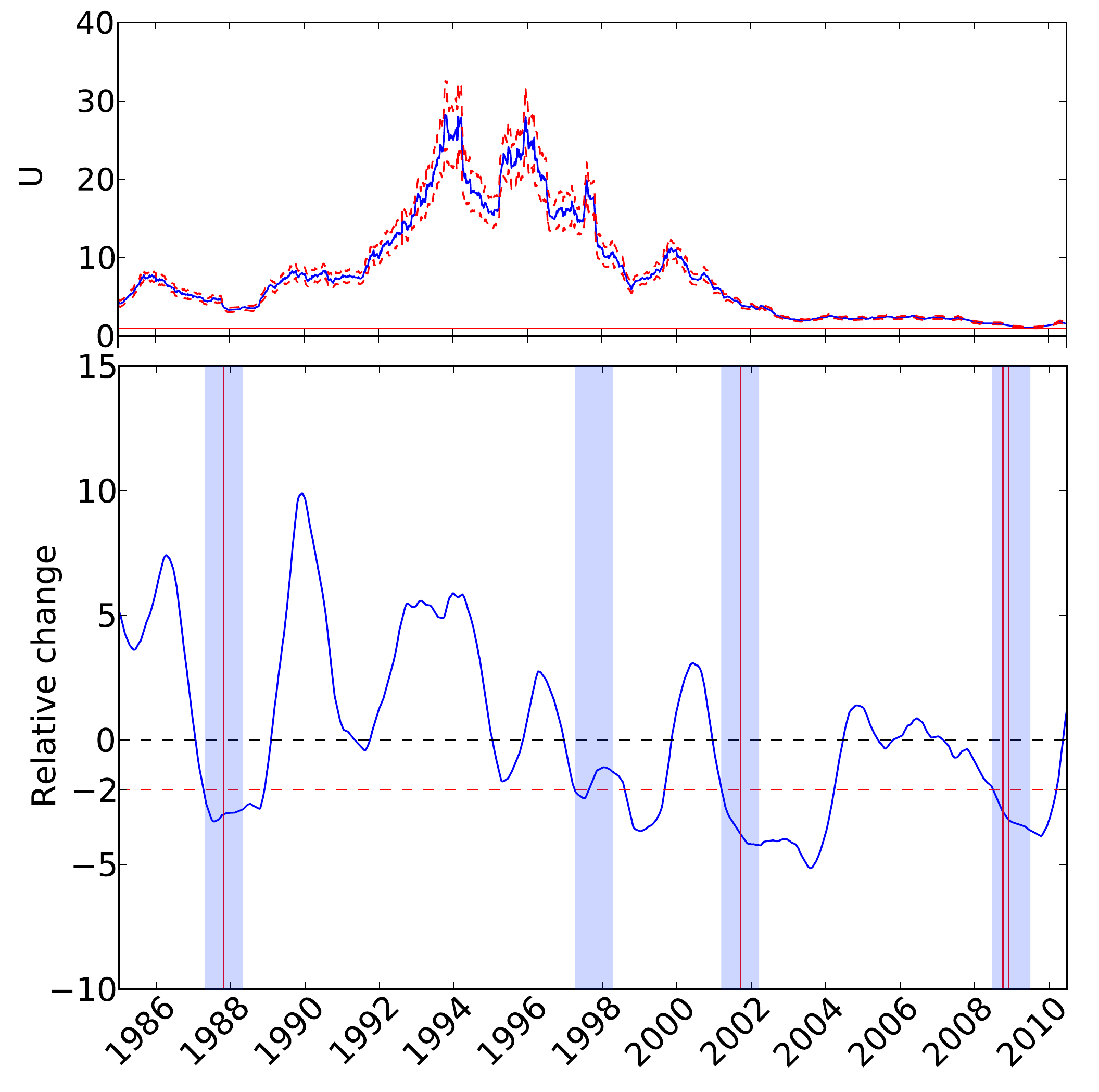}
\caption{Top panel is the same as in Figure~\ref{fig:parameter-short} for the period
  1985-2010. Bottom panel is the annual change of $\mathsf U$ as a fraction of its
  standard deviation from one year earlier averaged over the previous
  year.  Of the twenty largest percentage drops of the Dow Jones
  Industrial Average, eight are in the displayed time period:
  10/19/1987, 10/26/1987, 10/27/1997, 9/17/2001, 9/29/2008, 10/9/2008,
  10/15/2008, 12/1/2008 (vertical red lines). Four year-long windows (shading)
  follow two standard deviation drops in the the model parameter after periods
  of increase. Positive-return distributions are computed from the
  daily returns of stocks of the Russell~3000 for dates after July 1,
  1999.  Before July 1, 1999, returns were obtained for the
  Russell~3000 membership lists from 2001, 2004, and 2007. Each
  positive return fraction was computed with more than~140 stocks.}
  \label{fig:parameter}
\end{figure}
In order to evaluate more broadly the predictive ability of the model,
we consider the period from 1985--2010
(Figure~\ref{fig:parameter}). While there was no other financial crisis
of comparable magnitude to the current one, drops in the model
parameter $U$ anticipate large drops of the Dow Jones Industrial Average
(DJI). The bottom panel of Figure~\ref{fig:parameter} shows the (annual) change in the model
parameter as a fraction of the standard deviation one year earlier,
averaged over the preceding year. Of the twenty largest percentage
drops of the DJI, eight are in the displayed time period
\cite{djia:drops}, proximate to Black Monday \cite{carlson}, the Asian
market crisis \cite{berg}, September 11, 2001, and the recent financial
crisis. A simple signature pattern precedes the drops by less than a
year:  after a period of positive change, a large drop occurs in the
parameter $U$, greater than twice the standard deviation from one year
earlier, when averaged over the preceding year. This pattern
identifies four year-long windows in which occur the eight largest
percentage drops of the DJI within the last~26 years ($p<0.00007$ for
four non-overlapping, year-long windows).

The performance of the predictive pattern is exceptional.  Two
questions might be asked to evaluate its robustness. First, the pattern is nearly matched in~1995
when the change of the parameter as a fraction of the standard
deviation drops to below~$-1.67$ in April, 1995, but this near match is
not followed by a large drop in the DJI within the year.  Secondly, the
drop in the DJI on September 17, 2001, on the trading day immediately
following September 11, 2001, appears to have a direct external cause,
and therefore we might not consider the intrinsic stability of the market as
predictive, though we do not exclude the possibility that a drop would have
occurred without the attack.  If we interpret the results conservatively, we
would eliminate the year 2001 from consideration ($p<0.0005$), include the
near-prediction in~1995 ($p<0.0004$), or both ($p<0.002$).  Even in
this case, there is strong predictive success.

Our prediction of the event on September 17, 2001 was also obtained by
Hurst time series analysis \cite{grech}, and our work provides
additional evidence that this event was not solely a reaction to the
events of September 11, but largely reflected intrinsic market
dynamics. On the other hand we do not predict an event for 2003. This
is to be contrasted with the predictions by others that did not come
true \cite{sornette2}. However, we do find a significant drop in $U$
prior to that time, suggesting increased vulnerability. It appears that
two events conspire to prevent the crash. First, the increase in
mimicry leveled off before the systemic instability threshold.
Moreover, following the smaller crash on September 17, 2001 there was
no actual recovery of the market dynamics, which continued to be
vulnerable, but without a crash, until 2007. Our result that increased
mimicry {\it anticipates} panics is also distinct from debates about
the origins of higher correlations that {\it follow} crises \cite{longin,forbes,caporale}.

Central to the discussion of panic in the literature
\cite{wolfenstein,smelser,quarantelli,mawson} is the degree to
which it reflects external threats that cause each individual to panic,
or whether it reflects mimicry with or without external causes. Even
when mimicry is important, underlying conditions that imply increased
risk can elevate sensitivity and the tendency to mimicry. Underlying
conditions in this context may include internal trends such as market
bubbles or external factors such as war, or the financial disruptions
that preceded the recent market decline.  When panic involves
collective action, rather than individual response, precursor
fluctuations are likely to exist due to a growing sensitivity to real
or random disturbances.
Our results suggest that self-induced panic is a critical component of
both the current financial crisis and large single day drops over
recent years. The signature we found, the existence of a large
probability of co-movement of stocks on any given day, is a measure of
systemic risk and vulnerability to self-induced panic. Finally, we note
that the ability to distinguish between self-induced panic and the
result of external effects may be widely applicable to collective
behaviors \cite{bar-yam}.

Finally, we note that since volatility is often considered a measure of risk, its magnitude might be considered a more natural predictor of crises than the measure of collective following we have presented. Although volatility does increase at the onset of a crisis, and may remain elevated, it provides an unreliable crisis predictor of market crashes. Detailed comparison between our method and volatility will be discussed in a forthcoming paper.\cite{volatility}

Acknowledgements: We thank James H. Stock, Jeffrey C. Fuhrer and
Richard Cooper for helpful comments. I.R.E. thanks the Radcliffe Institute
for a fellowship. M.A.M.A. thanks FAPESP and CNPq for financial support.

\begin{appendix}
 
\section{Dynamic network model of daily stock returns}

\subsection{Model}
\label{subsec:model}

Consider a network representing an economic market with $N$ variable
nodes taking only the values $-1$ or $1$, representing decreasing or
increasing returns of a particular stock. In addition there are $D$
and $U$ nodes frozen in state $-1$ and $1$ respectively. At each time
step a variable node is selected at random; with probability $1-p$ the
node copies the state of one of its connected neighbors, and with
probability $p$ the state remains unchanged. The frozen nodes are
interpreted as external perturbations with negative and positive
effects on the returns. Analytically extending $D$ and $U$ to be real
numbers enables modeling arbitrary strengths of external
perturbations. \cite{chinellato}

For a fully connected network the behavior of the system can be solved
exactly as follows. The nodes are indistinguishable and the state of
the network is fully specified by the number of nodes with internal
state $1$. Therefore, there are only $N+1$ distinguishable global
states, which we denote $\sigma_k$, $k=0,1,\ldots,N$. The state
$\sigma_k$ has $k$ variable nodes in state $1$ and $N-k$ variable
nodes in state $-1$. If $P_t(m)$ is the probability of finding the
network in the state $\sigma_m$ at the time $t$, then $P_{t+1}(m)$ can
depend only on $P_t(m)$, $P_t(m+1)$ and $P_t(m-1)$. The probabilities
$P_t(m)$ define a vector of $N+1$ components ${\bf P}_t$. In terms of
${\bf P}_t$ the dynamics is described by the equation
\begin{displaymath}
{\bf P}_{t+1} = {\bf T} {\bf P}_t \equiv \left( {\bf 1} -
\frac{(1-p)}{N(N+D+U-1)} {\bf A}\right) {\bf P}_t \label{timev}
\end{displaymath}
where the time evolution matrix ${\bf T}$, and also the auxiliary
matrix ${\bf A}$, is tri-diagonal. The non-zero elements of ${\bf
A}$ are independent of $p$ and are given by
\begin{displaymath}
\begin{array}{l}
A_{m,m} = 2m(N-m) + U(N-m) + D m \\
A_{m,m+1} = -(m+1)(N+D-m-1) \\
A_{m,m-1} = -(N-m+1)(U+m-1).
\end{array}
\end{displaymath}
The transition probability from state $\sigma_M$ to $\sigma_L$ after a
time $t$ can be written as
\begin{displaymath}
P(L,t;M,0) = \sum_{r=0}^N  b_{rM} a_{rL} \lambda_r^t \;.
\end{displaymath}
where $a_{rL}$ and $b_{rM}$ are the components of the right and left
$r$-th eigenvectors of the evolution matrix, ${\bf a}_r$ and ${\bf b}_r$.
Thus, the dynamical problem has been reduced to finding the right and
left eigenvectors and the eigenvalues of ${\bf T}$.

The eigenvalues $\lambda_r$ of ${\bf T}$ are given by
\begin{displaymath}
\lambda_r = 1- \frac{(1-p)}{N(N+D+U-1)}r(r-1+D+U)
\end{displaymath}
and satisfy $0 \leq p \leq \lambda_r \leq 1$. The equation for
$P(L,t;M,0)$ shows that the asymptotic state of the network is
determined only by the right and left eigenvectors with unit
eigenvalue, i.e., by the eigenvectors of $\lambda_0=1$. The
coefficients of the corresponding (unnormalized) left eigenvector are
simply $b_{0k}=1$. The coefficients $a_{0k}$ of the right eigenvector
are given by the Taylor expansion of the hypergeometric function
$F(-N,U,1-N-D,x) \equiv \sum_k a_{0k} x^k$.  After normalization
these coefficients give the stationary distribution
\begin{equation}
\rho(k)=\frac{
{U+k-1\choose k}
{N+D-k-1 \choose N-k}}{
{N+D+U-1 \choose N}}.
\label{eqn:rho}
\end{equation}
This is the probability of finding the network with $k$ nodes $1$ in
equilibrium and it is independent of the initial state. The other
eigenvectors can also be calculated and are also related to
hypergeometric functions.

\begin{figure}
  \sf
  \centering
  \includegraphics[height=4in]{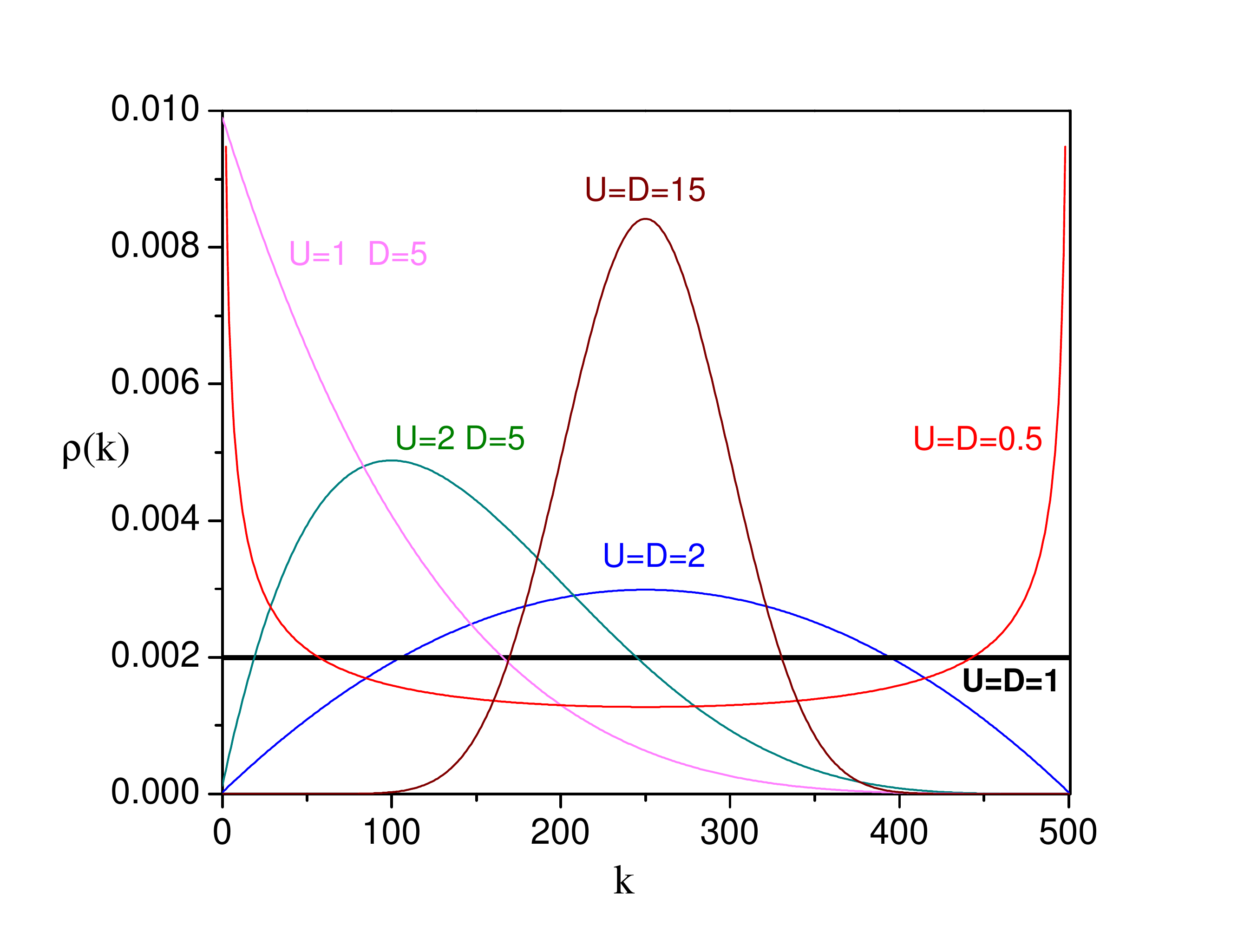}
  \caption{Probability distributions of finding the network with $k$ nodes
$1$ in
    equilibrium for different values of $U$ and $D$. The number of
variable nodes is $N=500$}
\label{figsom}
\end{figure}

One important feature of this solution is that for $U = D = 1$ we
obtain $\rho(m) = 1/(N + 1)$ for all values of N, i.e., U=D=1 is the
critical value of this model.  Thus all states $\sigma_k$ are equally
likely and the system executes a random walk through the state space.
For $U,  ~D >> 1$ $\rho(k)$ resembles a Gaussian distribution, but its
shape is very different for $U,  ~D << 1$, as illustrated in Figure
\ref{figsom} for $N=500$.

It might seem that the
critical point should depend on the size of the external influence
relative to the number of nodes in the system, i.e. $U/N$. However,
this is an order to disorder transition, and, as with the temperature
in physics models of phase transitions, the critical value does not
depend on the system size. For all values of $U=D$, the nodes have
equal probability of being $+$ or $-$. Thus, each node experiences an
environment that drives it equally toward positive and negative values.
The role of the external influence is only as a perturbation promoting
the transitions between states of the distribution. In this context,
even though the external influence on any one node decreases as $N$
increases, the influence across all nodes is independent of $N$. This
is because each node picks the external node to copy in proportion to
$1/N$. Thus, the average number of nodes that are changed per time step
by the external influence is independent of $N$.

This model has an analogue in population genetics and can be mapped
exactly into the Wright-Fisher-Moran model with two alleles and
mutation.\cite{ewens}
Consider a population of $N$ haploid individuals and a gene with
alleles $A_1$ and $A_2$. Sexual reproduction occurs between random
pairs of individuals with the offspring replacing one of the expiring
parents. After the allele of the offspring is chosen with equal
probability between the parents, there is also a probability $\mu_1$ to
mutate from $A_1$ to $A_2$ or $\mu_2$ to mutate from $A_2$ to $A_1$.
The number of alleles $A_1$ in the population in equilibrium is given
by Equation \ref{eqn:rho} with
\begin{equation}
U = \frac{2\mu_2(N-1)}{1-\mu_1-\mu_2} \qquad
D = \frac{2\mu_1(N-1)}{1-\mu_1-\mu2}.
\end{equation}
This problem was first considered by Watterson and Gladstein
\cite{watter61,glad78} with no mutation and latter generalized by
Cannings \cite{cann74}. A detailed account is given by Ewens \cite{ewens}.

For networks with different topologies the effect of the frozen nodes
is amplified. To see this we note that the probability that a variable
node
copies a frozen node is $P_i=(D+U)/(D+U+k_i)$ where
$k_i$ is the degree of the node. For fully connected networks $k_i=N-1$
and $P_i\equiv P_{FC}$. For general networks an average value $P_{av}$
can be calculated by replacing $k_i$ by the average degree
$k_{av}$. We
can then define effective numbers of frozen nodes, $D_{ef}$ and
$U_{ef}$, as being the values of $D$ and $U$ in $P_{FC}$ for
which $P_{av} \equiv P_{FC}$. This leads to
\begin{displaymath}
D_{ef} = f D, \qquad \qquad U_{ef} = f U
\label{eq:rescaling}
\end{displaymath}
where $f=(N-1)/k_{av}$. Therefore, as the network acquires more
internal connections and $k_{av}$ increases, the effective values of
$D$ and $U$ decrease.

\subsection{Curve fits}

Theoretical fits are computed from an unbiased estimator of the
standard deviation.  The distribution described in
Section~\ref{subsec:model} takes values $k=0,\ldots,N$.  We are
interested in the positive fraction, or $k/N$, rather than the number
of positive nodes.  The central moments of the positive fraction
distribution can be computed from Equation~\ref{eqn:rho}.  We express
the mean, $c_1$, and variance, $c_2$ in terms of $\xi=U/(U+D)$ and
$a=U+D$.
\begin{eqnarray}
c_1&=& \xi \label{eqn:c1}\\
\noalign{\medskip}
c_2&=& \frac{\xi(1-\xi)(1+a/N)}{a+1}\label{eqn:c2}
\end{eqnarray}
Equations~\ref{eqn:c1} and~\ref{eqn:c2} can be inverted to solve for
$\xi$ and $a$:
\begin{eqnarray*}
\xi&=&c_1\\
\noalign{\medskip}
a&=&\frac{c_1(1-c_1)-c_2}{c_2-c_1(1-c_1)/N}
\end{eqnarray*}
For the case of stocks, fits using $c_1=0.5$ are better fits as
measured by the $\chi^2$ goodness-of-fit test than fits achieved by
setting $c_1$ to the mean of the empirical distribution.

\subsection{Data sources}

To compute empirical distributions, we used daily returns from the
Russell~3000, restricted to stocks trading on the NYSE, NYSE
Alternext, Nasdaq Capital, and Nasdaq Stock markets.  The Russell~3000
is maintained by Russell Investments, and is reconstrustructed  every
twelve months, with the new composition announced near the end
of June.  The list includes the largest~3000 US stocks trading on
market exchanges by market capitalization.  Specific details of the
selection process may be obtained from Russell
Investments~\cite{russell-methodology}.

To compute the empirical distribution of the positive-return fraction, we
used two methods.  For the period from July 1999 to June 2010, we
retrieved daily returns of stocks from the Russell~3000 membership
lists published at the end of June for the years 1999 through 2009.
Daily returns of the stocks on the list were retrieved for the
following twelve month period beginning in July.  Stocks that were
delisted during this period were included for all days before
delisting.  For the period before July 1999, we combined ticker symbols
from the Russell~3000 membership lists from June 2001, 2004, and 2007,
and retrieved daily returns for the symbols back to 1985.  Each
positive return fraction was computed with more than~140 stocks.

\end{appendix}

\end{document}